\documentstyle[]{l-school}
\def\review#1, #2, 1#3#4#5, #6 {{\sl#1\/} {\bf#2} (1#3#4#5) #6 }

\newcommand{\be}{\begin{equation}}
\newcommand{\eq}{\end{equation}}

\begin{document}
\markboth{Burkardt and El-Khozondar}{A 3+1 DIMENSIONAL LF-MODEL WITH SPONTANEOUS
$\chi SB$}
%
\setcounter{part}{10}
%
\title{A 3+1 Dimensional LF-Model with\\ Spontaneous $\chi SB$}
%
\author{M. Burkardt and H. El-Khozondar}
%
\institute{Department of Physics\\
New Mexico State University\\
Las Cruces, NM 88003-0001, U.S.A.
}
\maketitle
%
%
%
%
%
\section{INTRODUCTION}
Light-front (LF) quantization provides an intuitive (physical basis!)
description of hadron structure that stays close to the relevant degrees of
freedom in high energy scattering: many high-energy scattering processes
probe hadrons along a light-like direction, since  
particles at very high energies travel close to the light-cone.
More detailed discussions 
of this main motivation for studying LF field theories can be found in 
Refs. \cite{mb:adv,bigguy}
as well as in the lecture by Stan Brodsky \cite{stan}. 

If the main application for LF field theory is supposed to be the phenomenology
of high-energy scattering processes, must one worry about the structure
the vacuum in the LF formalism? At first one might think that the answer to this
question is no. However, since the naive (no 0-modes) LF vacuum is known to
be trivial, one might worry, for example, whether deep-inelastic structure
functions can be correctly calculated on the
LF in a theory like QCD where the vacuum is known to have a nontrivial structure
and where one knows that this nontrivial vacuum structure plays an important
role for phenomenology.

It is well known that LF Hamiltonians allow for a richer 
counter-term structure \cite{osu:all}, and
spontaneous symmetry breaking in
normal coordinates can manifest itself as
explicit symmetry breaking counter-terms
in the corresponding LF Hamiltonian. In other words, the 
vacuum structure is shifted from states to fields.
Thus, one can account for
a nontrivial vacuum structure in the renormalization procedure.
Some immediate questions that arise in 
this context are
\begin{itemize}
\item Can a LF Hamiltonian, with a trivial vacuum,
have the same ``physics'' (in the sense of physical spectrum or
deep inelastic structure function) as an equal time Hamiltonian
with nontrivial vacuum?
\item What are implications for renormalization, i.e. how does one
have to renormalize in order to obtain the same physics?
\item What is the structure of the effective interaction for 
non-zero-modes
\end{itemize}
Of course, the general answer (i.e. $QCD_{3+1}$) is difficult to find,
but above questions have been studied in simple examples
\footnote{For a
more complete list of examples and
references on this topic, see Ref. \cite{bigguy}.} :

$QED/QCD_{1+1}$ \cite{eps,ho:vac,zhit}, Yukawa$_{1+1}$ \cite{mb:parity}, 
scalar theories (in any number of dimensions)
\cite{fr:eps}, perturbative
$QED/QCD_{3+1}$ \cite{mb:alex} and
``mean field models'': Gross-Neveu/NJL-model \cite{njl}.

The goal of such toy model studies  is to
build intuition which one can hopefully apply to $QCD_{3+1}$ 
(using trial and error).
However, while these models have been very useful for studying nonperturbative
renormalization in 1+1 dimensional LF field theories, it is not clear
to what extend these results can be generalized to 
sufficiently nontrivial theories in 3+1 dimensions.

\section{A 3+1 DIMENSIONAL TOY MODEL}

One would like to study a 3+1 dimensional model which goes
beyond the mean field approximation (NJL !), but on the other hand
being too ambitious results in very difficult or unsolvable models.
\footnote{For example, demanding Lorentz invariance, chiral symmetry and 
asymptotic freedom leaves QCD as the most simple model.}
We decided to place the following 
constraints on our model: 
\begin{itemize}
\item Most importantly, the model should be 3+1 dimensional,
but we do not require full rotational invariance.
\item The model should have spontaneous $\chi$SB  (but not just mean field)
\item Finally, it should be solvable both on the LF and using a conventional
technique (to provide a reference calculation).
\end{itemize}

Given these constraints, the
most simple model that we found is described by the Lagrangian
\begin{eqnarray}
{\cal L} = \bar{\psi_k}\left[ \delta^{kl}\left(i\partial\!\!\!\!\!\!\not \;\; -
m  \right)
-\frac{g}{\sqrt{N_c}}{\vec \gamma}_\perp {\vec A}^{kl}_\perp \right]\psi_l -
\frac{1}{2}
{\vec A}^{kl}_\perp \left(\Box +\lambda^2\right){\vec A}^{kl}_\perp ,
\label{eq:lager}
\end{eqnarray}
where $k,l$ are ``color'' indices ($N_c\rightarrow \infty$),
$\perp =x,y$ and where a
cutoff is imposed on the transverse momenta. A fermion mass was introduced
to avoid pathologies associated with the strict $m=0$ case.
$\chi SB$ can be studied by considering the $m\rightarrow 0$ limit of the
model.

The reasons for this bizarre choice of model [Eq. (\ref{eq:lager})] 
are as follows.
If one wants to study spontaneous breaking of chiral symmetry, then
one needs to have a chirally invariant interaction to start with, which
motivates a vector coupling between fermions and bosons. However, we
restricted the vector coupling to the $\perp$ component of a vector field
since otherwise one has to deal with couplings to the bad current
$j^-$ \footnote{$j^-$ is bilinear in the constrained component of the
fermion field, which makes it very difficult to renormalize this component
of the current in the LF framework.}. 
In a gauge theory, such couplings can be avoided by choice of gauge, but
we preferred not to work with a gauge theory, since this would give rise
to additional complications from infrared divergences.
Furthermore, we used a model with ``color'' degrees of freedom and
considered the limit where the number of colors is infinite, because such
a model is solvable, both on and off the LF. No interaction among the bosons
was included because this would complicate the model too much.
Finally, we used a cutoff on the transverse momenta because such a cutoff can
be used both on the LF as well as in normal coordinates and therefore
one can compare results from these two frameworks already for finite
values of the cutoff.  

\section{DYSON-SCHWINGER SOLUTION OF THE MODEL}
\begin{figure}
\unitlength1.cm
\begin{picture}(14,3.5)(.5,-12)
\includegraphics{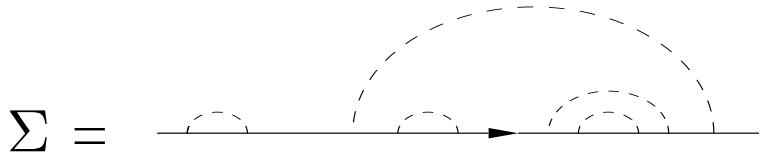}
\end{picture}
\caption{Typical Feynman diagram contributing to the fermion self-energy
in the large $N_C$ limit of the model. 
No crossed ``gluon'' lines are allowed.}
\label{fig:rain}
\end{figure}

Because we are considering the limit 
$N_C \rightarrow \infty$, of Eq. (\ref{eq:lager}), the iterated
rainbow approximation (Fig. \ref{fig:rain})
for the fermion self-energy $\Sigma$ becomes exact, yielding
\begin{eqnarray}
\Sigma(p^\mu) &=& ig^2 \int \frac{d^4k}{(2\pi )^4} {\vec \gamma}_\perp
S_F(p^\mu-k^\mu){\vec \gamma}_\perp \frac{1}{k^2-\lambda^2+i\varepsilon}
\nonumber\\
&=& \not \! \! p_L\Sigma_L({\vec p}_L^2,{\vec p}_\perp^2) + 
\Sigma_0({\vec p}_L^2,{\vec p}_\perp^2),
\label{eq:ds}
\end{eqnarray}
with:
\begin{equation}
S_F^{-1} = \not \! \! p_L\left[1-\Sigma_L({\vec p}_L^2,{\vec p}_\perp^2) \right]
+ \not \! \! p_\perp - \left[m+\Sigma_0({\vec p}_L^2,{\vec p}_\perp^2)\right] .
\end{equation}
These equations can be solved by iteration. From the self-consistently
obtained solution of the Dyson-Schwinger (DS) equation (\ref{eq:ds}) one can 
extract the
physical mass of the fermion. For sufficiently large coupling constant,
the physical mass for the fermion remains finite in the limit
$m \rightarrow 0$, proving the spontaneous breakdown of chiral symmetry
in the model.

\section{\bf LF-SOLUTION OF THE MODEL}
Since we wanted to investigate the applicability of the effective LF 
Hamiltonian formalism, we formulated above model without explicit
zero-mode degrees of freedom. In principle, the calculation should thus be
straightforward, using standard numerical techniques, such as DLCQ
\cite{pa:dlcq}. However, in this approach it is hard to take full advantage
of the large $N_C$ limit so that it is difficult 
to compare the obtained spectrum with the results from solving the DS
equation. Instead, we use the following 2-step procedure to obtain
a formal solution for the LF formulation
\begin{itemize}
\item[1.] First, we derive a self-consistent Green's function equation
which is equivalent
to the DLCQ calculation. The Green's function calculation was
originally derived by starting from the covariant calculation and performing
$k^-$ integrations first (throwing away zero modes in $k^+$). 
In order to convince even the skeptics
that this procedure is equivalent to DLCQ, 
we demonstrate numerically that,
for finite and fixed DLCQ parameter $K$, the spectrum obtained
by diagonalizing the DLCQ matrix and the spectrum obtained
by solving the Green's function equation self-consistently
\footnote{Replacing integrals by finite sums in order to account for
the finite DLCQ parameter $K$.}
are identical.
\item[2.] In the next step we compare the self-consistent Green's function 
equation with the DS equation. In order to facilitate the comparison with 
the LF calculation, we
rewrite the DS equation (\ref{eq:ds}), using a spectral representation
for the fermion propagator $S_F$.
In the resulting DS equation with the spectral density,
we  combine energy denominators, using Feynman parameter integral
and perform the longitudinal momentum integral covariantly.
\end{itemize}
Details of this procedure can be found in Ref. \cite{mb:hala}.
The main results from the comparison between LF and DS equations are as 
follows
\begin{itemize}
\item The LF Green's function equation and the DS equation are identical 
(and thus have identical solutions) if and only if
one introduces an additional (in addition to the self-induced inertias)
counterterm to the kinetic mass term for the fermion.
\item For fixed transverse momentum cutoff, this additional kinetic
mass term is finite.
\item The value of the vertex mass in the LF Hamiltonian is the same as 
the value of the current mass in the DS equation.
\item In the chiral limit, mass generation for the (physical)
fermion occurs through
the kinetic mass counter term
\end{itemize}

\section{\bf IMPLICATIONS FOR RENORMALIZATION}
We have studied a 3+1 dimensional model with spontaneous breaking of chiral
symmetry both in a LF framework as well as in a Dyson-Schwinger framework.
Our work presents an explicit 3+1 dimensional example demonstrating
that there is no
conflict between chiral symmetry breaking and trivial LF vacua
provided the renormalization is properly done.

The effective interaction
(after integrating out 0-modes) can be summarized by a
few simple terms --- which are already present in the canonical 
Hamiltonian.
The current quark mass in the covariant formulation and the ``vertex mass''
in the LF formulation
are the same if one does not truncate the Fock space and if one
uses the same cutoff on and off the LF.
This is perhaps surprising, since the vertex mass multiplies
the only term in the canonical Hamiltonian which explicitly breaks
(LF-) chiral symmetry. Thus one might think
that chiral symmetry breaking would manifest itself through a
nonzero vertex mass. If one does not truncate Fock space,
this is \underline{not} what happens in this model!
\footnote{Further studies show that
a renormalization of the vertex mass arises from a Tamm-Dancoff 
truncation but not from integrating out zero-modes \cite{mb:meson}.}
$\chi$SB, in the sense of physical mass generation for the fermion,
manifests itself through a ``kinetic mass'' counterterm.

Even though we determined the kinetic mass counter term by directly
comparing the LF and DS calculation, several methods are conceivable which 
avoid reference to a non-LF calculation in order to set up the LF problem.
One possible procedure would be to impose parity invariance for physical
observables as a constraint \cite{mb:parity}.

M.B. would like to acknowledge Michael Frank and Craig Roberts for 
helpful discussions on the Schwinger-Dyson solution to the model.
We would like to thank Brett vande Sande for carefully reading the manuscript.
This work was supported by the D.O.E. under contract DE-FG03-96ER40965
and in part by TJNAF.

%
%

\begin{thebibliography}{100}
\bibitem{mb:adv} M. Burkardt, Advances\ Nucl.\ Phys.\ {\bf 23}, 1 (1996).
\bibitem{bigguy} S. Brodsky, H-C Pauli and S. Pinsky, submitted to
Physics Reports, hep-ph/9705477. 
\bibitem{stan} S.J. Brodsky, these proceedings, hep-ph/9706236.
\bibitem{osu:all}
K. G. Wilson et al., Phys.\ Rev.\ D\ {\bf 49}, 6720 (1994).\bibitem{thooft} G.~'t~Hooft, Nucl.\ Phys. B\ {\bf 75}, 461 (1974).
\bibitem{eps} F.~Lenz et. al., A..\ Phys. {\bf 208}, 1 (1990).
\bibitem{ho:vac} K.~Hornbostel,
Phys.\ Rev.\ D\ {\bf 45} (1992) 3781;\\
D.G.~Robertson, Phys.\ Rev.\ D\ {\bf 47}, 2549 (1993).
\bibitem{zhit} A.~Zhitnitsky, Phys.\ Lett.\ B\ {\bf 165 }, 405 (1985);\\
M.~Burkardt, Phys.\ Rev.\ D\ {\bf 53}, 933 (1996).
\bibitem{mb:parity} M. Burkardt, Phys.\ Rev.\ D\ {\bf 54}, 2913 (1996).
\bibitem{fr:eps} E. V. Prokhvatilov and V. A. Franke,
Sov. J. Nucl. Phys. {\bf 49} (1989) 688;
M. Burkardt, Phys.\ Rev.\ D\ {\bf 47}, 4628 (1993);
E. V. Prokhvatilov, H. W. L. Naus and H.-J. Pirner,
Phys.\ Rev.\ D\ {\bf 51}, 2933 (1995); J.P.~Vary, T.J.~Fields and H.-J.~Pirner,
Phys.\ Rev.\
 D\ {\bf 53}, 7231 (1996).
\bibitem{mb:alex} M. Burkardt and A. Langnau, Phys.\ Rev.\ D\ {\bf 44},
3857 (1991).
\bibitem{njl} C. Dietmaier et al., Z.\ Phys.\ A\ {\bf 334}, 220 (1989);\\
K. Itakura and S. Maedan, Prog.\ Theor.\ Phys.\ {\bf 97}, 635 (1997). 
\bibitem{pa:dlcq} H.-C.~Pauli and S.~J.~Brodsky, Phys.\ Rev.\ D\ 
{\bf 32}, 1993 (1985); 
{\it ibid} 2001 (1985).
\bibitem{mb:hala} M. Burkardt and H. El-Khozondar,Phys.\ Rev.\ D\ 
{\bf 55}, 6514 (1997).
\bibitem{mb:meson} M. Burkardt, hep-ph/9705224.
\end{thebibliography}
\end{document}